\documentclass[aps,pre,reprint,showpacs,bibnotes,twoside,eqsecnum]{revtex4-1}
\usepackage{epsfig}
\usepackage{color}
\usepackage{bm}
\usepackage{latexsym}
\usepackage{amssymb}
\usepackage{float}
\usepackage{graphicx}
\usepackage{mathtools}
\usepackage{epstopdf}

\begin{document}
\newcommand{\hide}[1]{}
\newcommand{\tbox}[1]{\mbox{\tiny #1}}
\newcommand{\half}{\mbox{\small $\frac{1}{2}$}}
\newcommand{\sinc}{\mbox{sinc}}
\newcommand{\const}{\mbox{const}}
\newcommand{\trc}{\mbox{trace}}
\newcommand{\intt}{\int\!\!\!\!\int }
\newcommand{\ointt}{\int\!\!\!\!\int\!\!\!\!\!\circ\ }
\newcommand{\eexp}{\mbox{e}^}
\newcommand{\bra}{\left\langle}
\newcommand{\ket}{\right\rangle}
\newcommand{\EPS} {\mbox{\LARGE $\epsilon$}}
\newcommand{\ar}{\mathsf r}
\newcommand{\im}{\mbox{Im}}
\newcommand{\re}{\mbox{Re}}
\newcommand{\bmsf}[1]{\bm{\mathsf{#1}}}
\newcommand{\mpg}[2][1.0\hsize]{\begin{minipage}[b]{#1}{#2}\end{minipage}}

\title{Heat conduction in harmonic chains with L\'evy--type disorder}
\author{I.~F.~Herrera-Gonz\'alez}
\affiliation{Department of Engineering, Universidad Popular Aut\'onoma del Estado de Puebla, 21 Sur 1103, Barrio Santiago, Puebla, Pue., M\'exico}
\author{J.~A.~M\'endez-Berm\'udez }
\affiliation{
Departamento de Matem\'{a}tica Aplicada e Estat\'{i}stica, Instituto de Ci\^{e}ncias 
Matem\'{a}ticas e de Computa\c{c}\~{a}o, Universidade de S\~{a}o Paulo - Campus de S\~{a}o Carlos, 
Caixa Postal 668, 13560-970 S\~{a}o Carlos, SP, Brazil \\
and \\
Instituto de F\'isica, Benem\'erita Universidad Aut\'onoma de Puebla, Apartado postal J-48, Puebla 72570, M\'exico}

\date{\today}

\begin{abstract}
We consider heat transport in one-dimensional harmonic chains attached at its ends to Langevin heat baths. The harmonic chain has mass impurities where the separation $d$ between any two successive impurities is randomly distributed according to a power-law distribution $P(d)\sim 1/d^{\alpha+1}$, being $\alpha>0$. In the regime where the first moment of the distribution is well  defined ($1<\alpha<2$) the thermal conductivity $\kappa$ scales with the system size $N$ as $\kappa\sim N^{(\alpha-3)/\alpha}$ for fixed boundary conditions, whereas for free boundary conditions $\kappa\sim N^{(\alpha-1)/\alpha}$ if $N\gg1$. When $\alpha=2$, the inverse localization length $\lambda$ scales with the frequency $\omega$ as $\lambda\sim \omega^2 \ln \omega$ in the low frequency regime, due to the logarithmic correction, the size scaling law of the thermal conductivity acquires a non-closed form. When $\alpha>2$, the thermal conductivity scales as in the uncorrelated disorder case. The situation $\alpha<1$ is only analyzed numerically, where $\lambda(\omega)\sim \omega^{2-\alpha}$ which leads to the following asymptotic thermal conductivity: $\kappa \sim N^{-(\alpha+1)/(2-\alpha)}$ for fixed boundary conditions and $\kappa \sim N^{(1-\alpha)/(2-\alpha)}$ for free boundary conditions.     
\end{abstract}


\maketitle


\section{Introduction}
Understanding the statistical properties of open, many-particle systems is one of the challenges of nonequilibrium statistical mechanics. 
In particular, deriving the phenomenological laws from the microscopic dynamics is a current unsolved problem. For instance, one of such phenomenological equations is Fourier's law 
\begin{equation}
{\bf J}=-\kappa \nabla T,
\end{equation}
where ${\bf J}$ is the amount of heat transported through surface unit and per unit time, $T({\bf x},t)$ is the local temperature and $\kappa$ is the thermal conductivity. Since there is no general framework to explain the latter phenomenological law, one uses simple models to provide a more firm foundation to the heat conductivity and to understand more deeply the hypothesis behind it.

The simplest model that helps us to understand the statistical foundation of phenomenological equations is the ordered harmonic chain attached at its ends to two Langevin heath baths. In this case, heat transport is ballistic and the thermal conductivity scales with the system size $N$ as $\kappa \sim N$ \cite{CL71}, therefore, Fourier's law is not satisfied since $\kappa$ should be an intensive quantity. Moreover, the thermal conductivity diverges in the thermodynamic limit;~this phenomenon is known as anomalous heat conduction. One may think that the latter phenomenon appears due to the fact that there is no scattering of phonons in the isolated system, but this is not the case since even when one introduces phonon scattering by adding disorder or by including anharmonicity in the system the problem of anomalous heat conduction remains for the vast majority of the models (see \cite{D08,LLP03} and references therein). Even more, the anomalous heat conduction behavior predicted by some of these low dimensional lattices has been observed in carbon and boron-nitride nanotubes \cite{COGMZ08}, and in other quasi one-dimensional (1D) nanostructures \cite{LXXZZL12} which reveal that dimensionality plays a crucial role in determining the fundamental laws of the heat conduction.

When uncorrelated disorder is introduced in the harmonic chain, the phenomenon of Anderson localization emerges and normal modes become exponentially localized with the characteristic length $L_{\text{loc}}$, known as the localization length,  when $N\rightarrow \infty$; with the exception of the zero frequency mode that remains delocalized. Thus, when a finite size chain is connected to thermal baths, only the lowest frequency modes contribute to heat conduction. This prevents the system to be a thermal insulator, and produces a size scaling of the thermal conductivity of the form $\kappa\sim N^{-1/2}$ for fixed boundary conditions and $\kappa\sim N^{1/2}$ for free boundary conditions \cite{MI70,V71,RG71,V79}. In \cite{MI70,V71}, the harmonic chain is coupled at its ends to Langevin heat baths, while in \cite{RG71,V79}, oscillators heat baths are used instead. A unified description on the role of the spectral properties of  thermal baths to the heat conduction, where Langevin and oscillators baths are two special cases, is given in Ref. \cite{D01} where it is shown that one can even choose a heat bath with some peculiar spectral properties that leads to a finite thermal conductivity.

The role that correlations of the isotopic disorder plays in the heat conduction has recently attracted some attention for the following reasons: One can produce extended vibrational modes  within a frequency interval, in such situation $\kappa \sim N$ regardless on boundary conditions \cite{HIT10,MCRL03}. With the use of long range correlated disorder, it is possible to control the number of low frequency vibrational modes, and thus, the size scaling behavior of the thermal conductivity can be controlled too \cite{HIT15}. Short range correlated disorder plays an important role as well, some studies suggest that heat conduction in a disordered 1D lattice can be controlled
via statistical clustering of the constituent component atoms into domains \cite{OZ14}.

An interesting correlated disorder model has been proposed in Ref.~\cite{ZLW15}, where the localization properties  in a harmonic chain with mass impurities distributed according to a power law relation $P(d)\sim d^{-(1+\alpha)}$ is studied semi-analytically, being $d$ the distance between two successive impurities. This simple model has been introduced to provide a background of  wave properties of light (like interference) and to analyze the role of localization in L\'evy glasses, which has become a relevant issue in the last years due to the fact that these disordered materials can be engineered  to study L\'evy flights of light in a controlled way \cite{BBW08}.

In the harmonic chain model of Ref.~\cite{ZLW15}, the localization length can be controlled through the $\alpha$ parameter, this has lead to study electromagnetic wave propagation through multilayered structures with L\'evy--type disorder \cite{AN15,AN18} and to study heat conduction through a chain of harmonic springs whose spring constants are distributed randomly with a L\'evy--type distribution \cite{AOI18,AABOI19}.

In this work, we study localization properties and heat conduction in the correlated disorder model introduced in Ref.~\cite{ZLW15}. We derive a complete analytical solution for the localization length when $1<\alpha \le 2$. When $\alpha \le 1$, the localization length is only studied via numerical simulations. It is worthwhile to mention that  the cases $\alpha<1$ and $\alpha=2$  have not been previously studied. An analytical estimation for the asymptotic scaling law of the thermal conductivity with the system size is also obtained. 

\section{L\'evy--type disordered lattice: Statistical properties}
\label{two}
 We consider a one dimensional harmonic chain composed of $N$ masses $m_1$,...,$m_N$ with  nearest neighbors interactions having the same spring constant $k$ throughout the chain and a distance between two consecutive equilibrium positions denoted by $a$. The equation of motion for the mass $m_n$ is given by
 \begin{equation}
 m_n \ddot{u}_n=k(u_{n+1}-u_n)+k(u_{n-1}-u_n),
 \label{motion}
 \end{equation}
 where $u_{n}$ is the displacement of the $n$th mass from its equilibrium position $na$. For the moment, we do not specify boundary conditions since all the results in this and next sections are derived using the condition $N\rightarrow \infty$.
 
\begin{figure}[!t]
\includegraphics[scale=.42]{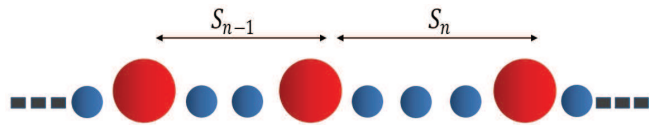}
\caption{(Color online) Schematic illustration of the model. Red large circles represent the defects of mass $M$, whereas the blue small ones depict the background masses which take the value $m$. Here, $S_{n-1}=3$ and $S_n=4$.}
\label{Fig0}
\end{figure}

The masses $m_n$ can only take two possible values: $m$ and $M$; where the separation between the equilibrium positions of any two consecutive masses $M$ is given by $sa$ with $s$ a natural number randomly distributed according to the Zeta distribution \cite{JKB93}
\begin{equation}
P(s)=\frac{s^{-(\alpha+1)}}{\zeta(\alpha+1)}, \ \ \alpha>0 ,
\label{distri}
\end{equation}
being $\zeta(z)$ the Riemann zeta function. The missing $s-1$ masses between any two successive masses $M$ acquire the value $m$  as Fig.~\ref{Fig0} shows.

The first moment of the Zeta distribution is not well defined when $0<\alpha \le1$ and diverges as 
\begin{eqnarray}
\langle s \rangle \sim \left\{
                \begin{array}{lcc}
                  N^{1-\alpha}   &\mbox{if}&  0<\alpha<1 \\
                  \ln N    &\mbox{if}&  \alpha=1.
                \end{array} 
         \right.
         \label{div}
\end{eqnarray}
Here $N\gg 1$ denotes the number of  addends  involved in the calculation of $\langle s \rangle$. Whereas, $\langle s \rangle= \zeta(\alpha)/\zeta(\alpha+1)$ is finite for $\alpha>1$, and the second moment of the distribution is only well defined for $\alpha>2$. Thus, $M$ represents the mass of the defects, while $m$ depicts the background mass. The number of defects $N_d=\zeta(\alpha+1)N/\zeta(\alpha)$, expression valid for $\alpha>1$, determines the random succession of masses $m_n$ with first moment and variance given by 
\begin{eqnarray}
\langle m_n \rangle&=& m\left(\frac{\zeta(\alpha)-\zeta(\alpha+1)}{\zeta(\alpha)} \right) \nonumber \\
\mbox{var}[m_n]&=&\frac{\zeta(\alpha)-\zeta(\alpha+1)}{\zeta^2(\alpha)} \zeta(\alpha+1) (M-m)^2.
\label{moments}
\end{eqnarray}
Here, we have use the ergodic hypothesis in order to replace the ``time average" $\sum^N_{i=1}/N$ by the disorder average $\langle \cdot \rangle$ in the thermodynamic limit. If $\alpha<1$, $\langle m_n \rangle=m$ and $\mbox{var}[m_n]=0$ due to the fact that the density of the defects $\rho=N_d/N$ vanishes in the thermodynamic limit.

\begin{figure}[!t]
\includegraphics[width=8.8cm,height=7.4cm]{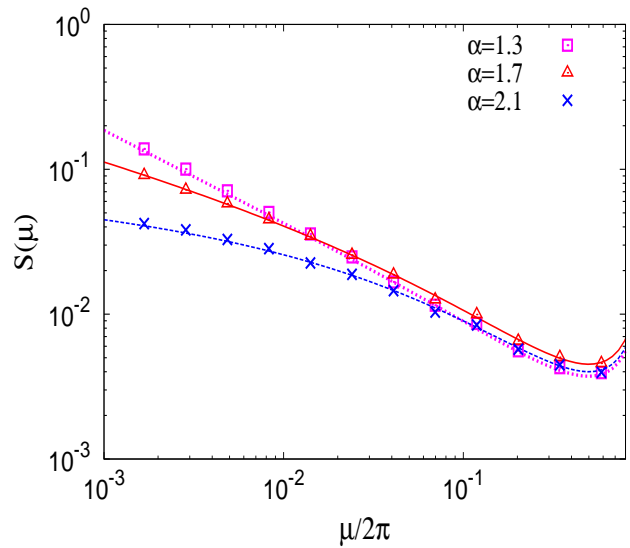}
\caption{(Color online) Power Spectrum as a function of the wave length for three different values of $\alpha$. Different point styles represent numerical data, while  different line styles are the analytical results of Eqs.~(\ref{char}) and (\ref{powert}). A system size of length $N=10^5$ is used for the numerical simulations, then $S(\mu)$ is averaged over $5\times10^4$ different realizations of the disorder. $M=0.3$ and $m=0.1$.}
\label{Fig1}
\end{figure}

One of the most important statistical properties in this work is the power spectrum $S(\mu)$ since the thermal properties of the harmonic chain will be determined by the low--wave-number $\mu$ behavior of $S$. It is convenient to write the power spectrum in its most general form
\begin{equation}
S(\mu)=\frac{1}{N}\left\langle \left|\sum^N_{n=1} m_n\exp(-ikn) \right|^2\right\rangle
\label{power}
\end{equation}
to facilitate both numerical and theoretical calculations. One can obtain an analytical result for the power spectrum for dichotomic disorder, explained in this section, if one rearrange the sum in Eq.~(\ref{power}) into several geometric series, therefore, after performing the disorder average and some algebra we get
\begin{eqnarray}
\label{powert}
S(\mu)&=&\frac{(M-m)^2}{\langle s \rangle}\frac{1-\left|\phi(\mu)\right|^2}{\left|1-\phi(\mu) \right|^2}, \\ 
\phi(\mu)&=&\left\langle \mbox{e}^{i\mu s}\right\rangle\equiv\frac{1}{\zeta(\alpha+1)} \sum^{\infty}_{s=1} \frac{\mbox{e}^{i\mu s}}{s^{\alpha+1}},
\label{dir}
\end{eqnarray}
where $\phi(\mu)$ is the characteristic function of the Zeta distribution which  can be written in terms of the polylogarithm $\mbox{Li}_{\alpha+1}(z)$ 
\begin{eqnarray*}
\phi(\mu)=\frac{\mbox{Li}_{\alpha+1}(\mbox{e}^{i\mu})}{\zeta(\alpha+1)}.
\end{eqnarray*}
With the process of analytic continuation, one can expand the definition of the polylogarithm to include complex values of $z$ for which $|z|>1$. Thus, the characteristic function for the Zeta distribution can be written in a more general form as a Bose-Einstein integral 
\begin{equation}
\phi(\mu)=\frac{1}{\Gamma(\alpha+1)\zeta(\alpha+1)}\int^{\infty}_0\frac{x^{\alpha}dx}{\mbox{e}^{t-i\mu}-1},
\label{char}
\end{equation}
where $\Gamma(z)$ is the Gamma function. Indeed, one can expand the integrand of Eq.~(\ref{char}) into power series, and after performing the integration term by term, the Dirichlet series of $\phi(\mu)$, given by Eq.~(\ref{dir}), is recovered. Expressions~(\ref{power}) and (\ref{char}) (or Eqs.~(\ref{power}) and (\ref{dir})) represent our analytical solution for the power spectrum which is in complete agreement with the numerical results as it is shown in Fig.~\ref{Fig1}. 

The polylogarithm can be represented as a Hankel contour integration which extends the Bose-Einstein integral representation to include a negative order parameter $\alpha+1$. Using such a representation one can find a series of the polylogarithm around $\mu=0$ \cite{W92}
\begin{equation}
\mbox{Li}_{\alpha+1}\left(e^{\mu}\right)=\Gamma(-\alpha)(-\mu)^{\alpha}+\sum^{\infty}_{n=1}\frac{\zeta(\alpha+1-n)}{n!}\mu^n,
\label{serie}
\end{equation}
an equation which is only valid for $|\mu|<2\pi$ and $\alpha\neq 0,1,2,3\dots$. If $\alpha+1$ is a natural number, the following  series representation must be used rather than Eq.~(\ref{serie}) \cite{W92}
\begin{equation}
\mbox{Li}_{\alpha+1}(\mbox{e}^{\mu})=\frac{\mu^\alpha}{\alpha!}[\mbox{H}_\alpha-\ln(-\mu)]+\sum^{\infty}_{n=0,n\neq\alpha}\frac{\zeta(\alpha+1-n)}{n!}\mu^n,
\label{serie1}
\end{equation}
where $H_n=\sum^n_{j=1} 1/j$ is the harmonic number and $H_0=0$. 

If one uses Eqs.~(\ref{serie}) or (\ref{serie1}), depending on whether $\alpha+1$ is a natural number or not, in Eq.~(\ref{powert}) and make the replacement of $\mu$ by $i\mu$ in the series representations of the polylogarithm about $\mu=0$, the low--wave-number behavior of the power spectrum is obtained ($\mu \rightarrow0$):  
\begin{eqnarray}
S(\mu)&=&\frac{-2(M-m)^2\zeta(\alpha+1)\cos(\frac{\pi}{2}\alpha)}{\langle s\rangle \Gamma(-\alpha)}\mu^{-\alpha},   \qquad  0<\alpha<1, \nonumber \\ 
S(\mu)&=&\frac{\pi^3(M-m)^2}{6\langle s\rangle}\frac{1}{\mu \ln^2 \mu}, \qquad \qquad \qquad \qquad \qquad  \alpha=1, \nonumber \\
S(\mu)&=&\frac{-2\zeta^2(\alpha+1)(M-m)^2 \Gamma(-\alpha)\cos(\frac{\pi}{2}\alpha)}{\zeta^3(\alpha)\mu^{\alpha-2}}, \ \  1<\alpha<2, \nonumber \\ 
S(\mu)&=&-\frac{216(M-m)^2\zeta^2(3)}{\pi^6}\ln \mu, \qquad \qquad \qquad \qquad \alpha=2. \nonumber \\                     
\label{powerT}
\end{eqnarray}
Thus, for $\alpha<1$, the power spectrum is size-dependent ($S\sim N^{\alpha-1}$, see Eq.~(\ref{div})) which means that in the thermodynamic limit $S(\mu)=0$, and if the system size is finite, $S \sim \mu^{-\alpha}$. When $1<\alpha<2$, the power spectrum also diverges when $\mu\rightarrow0$ as $S(\mu)\sim \mu^{\alpha-2}$. The latter results are in agreement with those presented in Ref.~\cite{ZLW15} which were obtained by numerical simulations. In addition, for $\alpha=2$, $S(\mu)\sim \ln \mu$ and for $\alpha>2$ the power spectrum tends towards a constant value, as $\mu \rightarrow0$, given by the following expression  
\begin{equation}
S(\mu)= \frac{\zeta(\alpha+1)\left[\zeta(\alpha-1)\zeta(\alpha+1)-\zeta^2(\alpha)\right](M-m)^2}{\zeta^3(\alpha)} .
\label{powerT1}
\end{equation} 
Finally, for $1<\alpha<2$, $S(\mu)$ has a power law singularity $\mu^{\alpha-2}$ in the low--wave-number regime, therefore, the disorder correlations $\langle m_n m_{n+l} \rangle$ must decay asymptotically ($l\gg1$) as a power law $|l|^{1-\alpha}$, which corresponds to the long-range correlations of the disorder.    
\section{Localization properties}
The localization length $L_{\text{loc}}$ which characterizes the scale over which the vibrational modes are exponentially localized is closely related to the rate of the exponential growth $\lambda$ of the discrete stochastic equation \cite{LLP03}  
\begin{equation}
R_{n+1}=2-\frac{m_n\omega^2}{k}-\frac{1}{R_n}, \quad \ R_n=\frac{v_n}{v_{n-1}},
\label{grow}
\end{equation}
where the quantity $v_n$ is defined by $u_n=v_ne^{i\omega t}$. When the latter relation is replaced in the equations of motion (\ref{motion}), Eq.~(\ref{grow}) is obtained. Thus, the explicit relation between both quantities is given by \cite{MI70}
\begin{eqnarray}
L^{-1}_{\text{loc}}=\lambda \equiv \lim_{N\rightarrow \infty}\frac{1}{N}\sum^N_{n=1} \ln \left(\frac{R_n}{R_{n-1}}\right) .
\label{rel}
\end{eqnarray}
$\lambda$ is a self averaged quantity  whose value does not depend on the initial condition $R_0$ and is known as the Lyapunov exponent. $L^{-1}_{\text{loc}}$  is measured in units of the lattice constant $a$. Equation (\ref{grow}) together with relation (\ref{rel}) are suitable both for theoretical and numerical calculations of the localization length. For numerical simulations, it has the advantage that one avoids computing overflow which occurs due to the exponential nature of the solution for $v_n$, so the technique of numerical renormalization for calculating the Lyapunov exponent is not needed.

There exists an explicit expression for the localization length  which is derived using the condition of weak disorder $\sqrt{\mbox{var}[m_n]}/\langle m_n \rangle\ll1$. Under such condition $L^{-1}_{\text{loc}}$ is given by \cite{HIT10}
\begin{eqnarray}
L^{-1}_{\text{loc}}(\mu)=\frac{\mbox{var}[m_n]}{2\langle m_n \rangle^2}\tan^2\left(\frac{\mu}{2}\right)W(\mu),
\label{locEPL}
\end{eqnarray}  
where
\begin{eqnarray}
W(\mu)=1+2\sum^{\infty}_{l=1}\chi(l)\cos(2l\mu)
\label{powern}
\end{eqnarray} 
is the power spectrum normalized to the variance of the random masses,  $\chi(l)$ is the normalized binary correlator $\chi(l)=(\langle m_nm_{n+l}\rangle -\langle m_n\rangle^2)/\mbox{var}[m_n]$, and the lattice constant $a$ is set to one. For weak disorder,  the frequency of the vibrational modes follows practically the same dispersion relation than in the ordered chain, then $\omega(\mu)=\omega_{\text{max}}\left|\sin\left(\mu/2\right)\right|$ with
\begin{eqnarray}
\omega_{\text{max}}=\sqrt{\frac{4k}{\langle m_n\rangle}},
\label{max} 
\end{eqnarray} 
being the largest frequency in the ordered chain and $\mu$ the wave number. Thus, the inverse localization length (\ref{locEPL}) can be written in terms of the frequency $\omega$ as well:
\begin{eqnarray}
L^{-1}_{\text{loc}}(\omega)=\frac{\mbox{var}[m_n]}{2\langle m_n \rangle^2}\frac{\omega^2}{\omega^2_{\text{max}}-\omega^2}W(\mu(\omega)),
\label{loc}
\end{eqnarray}
where $\mu(\omega)=2\arcsin \left(\left|\omega/ \omega_{\text{max}}\right|\right)$.

We can extend the validity range of Eq.~(\ref{loc}) using the following arguments: Notice that by taking the Fourier transform of the dynamical equations (\ref{motion}) with respect to time one obtains
\begin{equation}
u_{n+1}+u_{n-1}+4\left(\frac{\omega}{\omega_{\text{max}}}\right)^2\frac{\delta m_n}{\langle m_n \rangle}u_n=\left[2-4\frac{\omega^2}{\omega^2_{\text{max}}} \right]u_n,
\label{Fou}
\end{equation}
where $\delta m_n=m_n-\langle m_n \rangle$ is the fluctuation of the $n$th mass around its mean value and the term $V_e=4\left(\omega/\omega_{\text{max}}\right)^2\delta m_n/\langle m_n\rangle$ is an effective random potential felt by a mode of frequency $\omega$. Equation~(\ref{Fou}) has the same structure than the stationary Sch\"odinger equation for the 1D Anderson model with diagonal disorder for which an expression for the localization length is known for weak disorder \cite{IKM12} (see formula (5.23) of this reference); in our model the perturbative parameter is the effective potential $V_e$. Using the expression of the localization length for the 1D Anderson model and the analogy between both models, we get Eq.~(\ref{loc}) which is valid under the effective weak disorder condition
\begin{equation}
\left(\frac{2\omega}{\omega_{\text{max}}}\right)^2\frac{\sqrt{\mbox{var}[m_n]}}{\langle m_n \rangle}\ll1.
\label{weak}
\end{equation} 
This expression has a broader range of validity than the weak disorder assumption: On the one hand in the weak disorder condition, $\sqrt{\mbox{var}[m_n]}/\langle m_n \rangle\ll1$, Eq.~(\ref{loc}) works practically for the whole band frequency $[0,\omega_{\text{max}}]$, except for values of $\omega$ that are closed to the band edge $\omega_{\text{max}}$. On the other hand in the low frequency limit the condition of weak disorder can be removed and Eq.~(\ref{loc}) gets the form 
\begin{equation}
L^{-1}_{\text{loc}}(\omega)=\frac{\mbox{var}[m_n]}{2\langle m_n \rangle^2}\left(\frac{\omega}{\omega_{\text{max}}} \right)^2W\left(\frac{2\omega}{\omega_{\text{max}}}\right), \ \ \text{when} \ \ \omega \rightarrow 0.
\label{locw}
\end{equation}
The latter equation has also been derived in Refs.~\cite{HIT15,ZLW15}. 

Here, we must point out that the results derived for the localization length are only valid for disordered systems whose power spectrum is integrable in the interval $[0,\pi/2]$. Thus, for $\alpha\le1$ the power spectrum is no longer integrable and our analytical results can not be applied.  

Moreover, there is a relation between the power spectrum (\ref{power}) and the normalized power spectrum (\ref{powern}). Indeed, one can apply the stationary conditions of the random succession
$\{m_n \}$ in Eq.~(\ref{power}) and obtain the following relation: $W(\mu/2)=S(\mu)/\text{var}[m_n]$. Thus, using the dispersion relation, Eq.~(\ref{loc}) is written in terms of the power spectrum (\ref{power}), and therefore,
\begin{eqnarray}
\lambda(\omega)&=&\frac{\zeta(\alpha+1)\zeta(\alpha)}{A(\alpha)}\frac{\omega^2}{\omega^2_{\text{max}}-\omega^2} \frac{1-\left|\phi\left(2\mu(\omega)\right)\right|^2}{\left|1-\phi\left(2\mu(\omega)\right) \right|^2}, \nonumber \\
A(\alpha)&=&\left(\frac{m\left[\zeta(\alpha)-\zeta(\alpha+1)\right]+M\zeta(\alpha+1)}{M-m}\right)^2,
\label{locf}
\end{eqnarray}
being $\phi(\mu)$ and $\omega_{\text{max}}$ defined by eqs. (\ref{char}) and (\ref{max})-(\ref{moments}), respectively; $\mu(\omega)$ is defined through the dispersion relation $\mu(\omega)=2\arcsin \left(\left|\omega/ \omega_{\text{max}}\right|\right)$. 

Expressions (\ref{locf}) correspond to our analytical solution to the inverse localization length which is only valid under the effective weak disorder condition (\ref{weak}). In addition by
taking into account the low--wave-number behavior of $S(\mu)$  given in eqs. (\ref{powerT}) and (\ref{powerT1}), the low frequency limit of the inverse localization length is found:

For $1<\alpha<2$, and $\omega \rightarrow 0$
\begin{equation}
\lambda(\omega)=\frac{-2^{2\alpha-4}}{A(\alpha)\omega^{\alpha}_{\text{max}}} \frac{\zeta^2(\alpha+1)}{\zeta(\alpha)}\Gamma(-\alpha)\cos\left(\frac{\pi}{2}\alpha\right)\omega^\alpha.
\label{loc1}
\end{equation}

For $\alpha=2$, and $\omega \rightarrow 0$
\begin{equation}
\lambda(\omega)=\frac{3\zeta(3)\omega^2}{\pi^2 A(2)\omega^2_{\text{max}}} \left\lbrace \zeta(3)\left[\frac{3}{2}-\ln\left(\frac{4\omega}{\omega_{\text{max}}}\right)\right]-\frac{\pi^4}{36} \right\rbrace.
\label{loc2}
\end{equation}
In the latter equation, we take into account the next higher order term in the series expansion of $S(\mu)$ about $\mu=0$ which is a constant value. In this way, our numerical simulation matches the analytical result (\ref{loc2}) since the leading term $\ln \omega$ becomes only relevant for extremely low frequency values which are not available for numerical calculations.   

For $\alpha>2$, and $\omega \rightarrow 0$
\begin{equation}
\lambda(\omega)=\frac{\zeta(\alpha+1)}{2\zeta(\alpha)A(\alpha)}\left[\zeta(\alpha-1)\zeta(\alpha+1)-\zeta^2(\alpha) \right]\left(\frac{\omega}{\omega_{\text{max}}}\right)^2.
\label{loc3}
\end{equation}

\begin{figure}[!t]
\includegraphics[width=8.8cm,height=7.4cm]{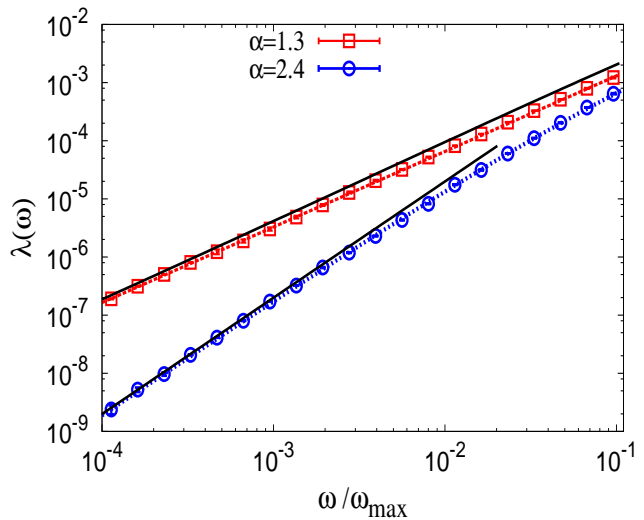}
\caption{(Color online) Inverse localization length as a function of the frequency for  different values of $\alpha$. Symbols are the numerical data, while dashed and dotted lines are the analytical result (\ref{loc}). The low frequency limit for $\lambda(\omega)$, given in Eqs.~(\ref{loc1}) and (\ref{loc3}), are shown in black solid lines. For the numerical simulations an average over $100$ different realizations of the disorder is taken into account and $k=1$.}
\label{Fig2}
\end{figure}

\begin{figure}[!t]
\includegraphics[width=8.8cm,height=7.4cm]{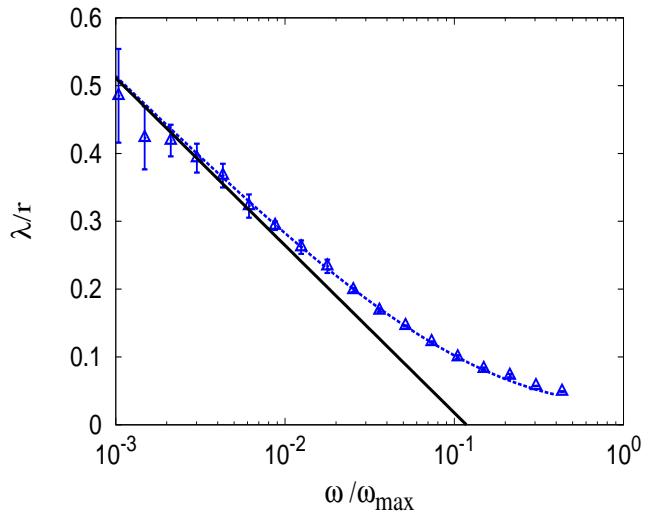}
\caption{(Color online) Inverse localization length, rescaled by the parameter $r=\left(\omega/\omega_{\text{max}} \right)^2$, as a function of the frequency. Triangles  are the numerical data, while dashed lines are the analytical result (\ref{loc}). The low frequency limit for $\lambda(\omega)$, given in Eq.~(\ref{loc2}), is shown as the black solid line. Same system size and ensemble average as in Fig.~\ref{Fig2}. $M=3$, $m=1$, and $k=1$.}
\label{Fig3}
\end{figure}

The comparison between analytical results (\ref{locf}), (\ref{loc1}), (\ref{loc2}) and (\ref{loc3}), and numerical data is done in Figs.~\ref{Fig2} and \ref{Fig3} which show an excellent agreement between both if $\omega$ is not too small and the general analytical result (\ref{locf}) is taken into account, whereas the corresponding low frequency regime is within the error bars. It is also seen in the figures that the inverse localization length given by Eq.~(\ref{locf}) tends towards its low frequency form predicted by Eqs.~(\ref{loc1}), (\ref{loc2}) and (\ref{loc3}). For the numerical simulation, we use Eqs.~(\ref{grow}) and (\ref{rel}) to compute the inverse localization length $L^{-1}_{\text{loc}}$ for a system of length $N=10^{11}$ (unless otherwise is stated), then a disorder average was performed over 10 disorder realizations. In this way, the time of numerical simulation was drastically reduced. Error bars are the standard deviation over the disorder realizations; the spring constant was set to one. 

\begin{figure}[!t]
\includegraphics[width=8.8cm,height=7.4cm]{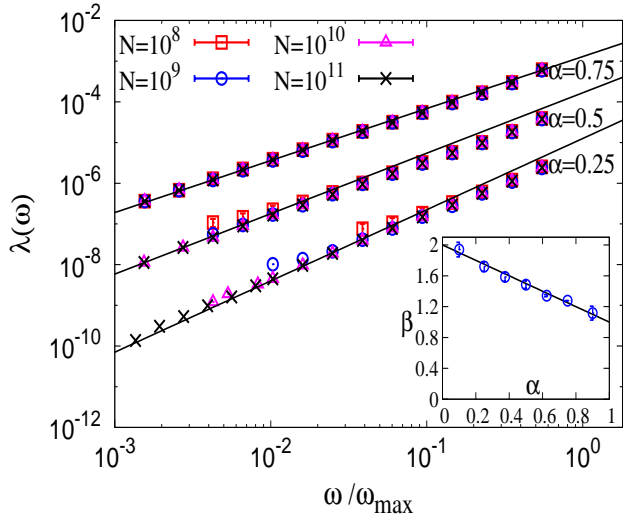}
\caption{(Color online) Inverse localization length as a function of the frequency for  different parameters $\alpha$ and system sizes $N$. An average over $100$ different realizations of disorder is taken into account, except for the case $N=10^{11}$, where the averaging is only performed over $10$ realizations of disorder. $M=0.3$, $m=0.1$ and $k=1$. Continuous lines correspond to the best fits of the low frequency numerical data to the functions $f(\omega)=c\omega^\beta$; for $\alpha=0.25$, $\beta=1.72 \pm 0.04$, for $\alpha=0.5$, $\beta=1.49 \pm 0.04$, and $\alpha=0.75$ gives $\beta=1.26\pm 0.01$. Other values of $\beta$ as a function of $\alpha$ are shown in the inset, where continuous black line depicts the best linear fit of these numerical data. Thus, the equation of this black line is $g(\alpha)=2.03\pm 0.04 -(1.02\pm 0.03)\alpha$; a system size of length $N=10^{11}$ is used for the latter fit.}
\label{Fig2N}
\end{figure}

In Ref.~\cite{ZLW15}, it is argued that the Lyapunov exponent vanishes in the thermodynamic limit for $\alpha <1$ due to the fact that the power spectrum goes to zero in this limit. However, our numerical results show that this is not the case (see Fig.~\ref{Fig2N}) since $\lambda$ results to be a size-independent quantity for sufficiently large system sizes, as usual. Moreover, the low frequency behavior of the Lyapunov exponent is
\begin{equation} 
\lambda(\omega)\sim \omega^{2-\alpha}, \qquad \qquad \alpha<1,
\label{am}
\end{equation}  
as the inset of the figure shows. Notice that  the same low frequency behavior is obtained if one assumes  that the power spectrum (\ref{powerT}) multiplied by $\omega^2$ is proportional to the localization length for $\alpha<1$, although this is not the case as it is explained above. From the same figure it is also seen that $\lambda(\omega) \rightarrow 0$ as $\alpha \rightarrow 0$, which is an expected result. 

From the above arguments, one may think that the low frequency behavior of the Lyapunov exponent would be proportional to $\omega/\ln^2\left(\omega/\omega_{\text{max}}\right)$ for $\alpha=1$, however, we try to prove this using numerical simulations, but no clear evidence of this scaling law is obtained even for system sizes of length $N=5\times 10^{11}$.  
 
\section{Thermal conductivity}
We now consider a finite harmonic chain composed by $N$ masses with the L\'evy--type disorder introduced in Sec.~\ref{two}. When the system is closed the equations of motion are given by (\ref{motion}). The chain is connected to two Langevin heat baths, thus the first and last masses of the chain are coupled to the heat baths with temperatures $T_1$ and $T_N$, respectively, and it is assumed that $T_1>T_N$. The equation of motion for the $n$th mass is given by the following Langevin equation
\begin{eqnarray}
m_n\ddot{u}_n&=&k(u_{n+1}+ u_{n-1}-2u_n)+\delta_{n,1}\left[\xi_1(t)-\gamma\dot{u}_1\right] \nonumber \\&+&\delta_{n,N}\left[\xi_N(t)-\gamma\dot{u}_N\right],
\label{Lmotion}
\end{eqnarray}
where $\xi_1$ and $\xi_N$ are Gaussian white noises corresponding to the left and right heat baths, respectively, and whose statistical properties are determined by the first moment and the autocorrelation function only,
\begin{eqnarray*}
\overline{\xi_j}&=&0 , \\
\overline{\xi_j(t)\xi_{j'}(t')}&=&2k_BT_j \gamma \delta_{jj'}\delta(t-t'),  \qquad \text{for} \ j=1,N.
\end{eqnarray*} 
Here $\overline{x(t)}$ denotes average over the stochastic process $x(t)$; in this way we can distinguish this average from the disorder average $\langle \cdot \rangle$. $\gamma$ represents the coupling strength of the baths with the harmonic chain and it has two effects in the equation of motions. On the one hand, it measures how strong the friction in the thermal baths is. On the other hand, it contributes to the intensity of the white noises; the relation between this two effects is given by the fluctuation dissipation theorem \cite{K78}.

Since now we are working with harmonic chains of finite size, a specification of the boundary conditions is needed. Two kinds of boundary condition are analyzed in this work: Free boundary conditions (free BC), which are determined by setting $u_0=u_1$ and $u_N=u_{N+1}$ in Eq.~(\ref{Lmotion}). This means that there is no external ``spring force" acting  on the harmonic chain. For fixed boundary conditions (fixed BC) the outer masses of the chain are coupled to springs of elastic constant $k$, not only to their nearest neighbors but also to external walls of infinite mass; therefore, $u_0=u_{N+1}=0$ in Eq.~(\ref{Lmotion}).
   
For discrete systems the thermal conductivity is defined through the expression \cite{LLP03}
\begin{eqnarray}
\kappa=\frac{J N}{\triangle T}, 
\label{ther}
\end{eqnarray} 
where $J$ and $\triangle T=T_1-T_N$ are the stationary heat flux through the harmonic chain and the temperature difference between the two thermal baths, respectively. In Ref.~\cite{HIT15} an analytical estimation has been done for the size scaling behavior of the thermal conductivity when the harmonic chain exhibits correlated mass disorder. When power spectrum of the fluctuations $\delta m_n$ of the random masses ($\delta m_n=m_n-\langle m_n\rangle$) behaves as $\mu^{\beta_1}$ in the low--wave-number limit, the thermal conductivity scales as
\begin{eqnarray}
    \kappa\sim
         \left\{
                \begin{array}{ll}
                  N^{(\beta_1-1)/(\beta_1+2)} & \ \mbox{for} \ \ \text{fixed BC} \\
                  N^{(\beta_1+1)/(\beta_1+2)} & \ \mbox{for} \ \ \text{free BC}
                \end{array}
         \right. .
\label{scale}
\end{eqnarray}

Due to the fact that the power spectrum of the random masses $m_n$ is equal to the corresponding one of the fluctuations $\delta m_n$, therefore, the $\beta$ exponent is determined from  the low--wave-number behavior of $S(\mu)$ given by Eqs.~(\ref{powerT}) and (\ref{powerT1}). In this way, and with the use of Eq.~(\ref{scale}), the scaling behavior of the thermal conductivity with system size is obtained in harmonic chains with  L\'evy--type  disorder: 

For $1<\alpha<2$
\begin{eqnarray}
    \kappa\sim
         \left\{
                \begin{array}{ll}
                  N^{(\alpha-3)/\alpha} & \ \mbox{for} \ \ \text{fixed BC} \\
                  N^{(\alpha-1)/\alpha} & \ \mbox{for} \ \ \text{free BC}
                \end{array}
         \right. .
\label{scaleT}
\end{eqnarray}

For $\alpha>2$
\begin{eqnarray}
    \kappa\sim
         \left\{
                \begin{array}{ll}
                  N^{-1/2} & \ \mbox{for} \ \ \text{fixed BC} \\
                  N^{1/2} & \ \mbox{for} \ \ \text{free BC}
                \end{array}
         \right. .
\label{scaleT1}
\end{eqnarray}
Notice that for free BC the exponent of the thermal conductivity becomes arbitrarily small as $\alpha \rightarrow 1$. Thus, an intensive thermal conductivity is not obtained neither for free BC nor for fixed BC. For $\alpha>2$, the asymptotic thermal conductivity   is equal to the size scaling law of the thermal conductivity in harmonic chains with uncorrelated disorder.

An special case appears when $\alpha=2$, for which the above results can not be applied since $S(\mu)\sim \ln \mu$ when $\mu \rightarrow 0$. For this reason, we use the following formulas derived in \cite{HIT15}:
\begin{eqnarray}
N_e\simeq \frac{2\langle m_n \rangle}{\pi}\sqrt{\frac{2N}{\text{var}[m_n]W\left(\pi N_e/N\right)}} , \label{for} \\
  J  \sim
         \left\{
                \begin{array}{ll}
                  \left(N_e/N\right)^3 & \ \mbox{for} \ \ \text{fixed BC} \\
                  N_e/N & \ \mbox{for} \ \ \text{free BC}
                \end{array}
         \right. ;
\label{for1}
\end{eqnarray}
where $N_e$ is the number of extended low frequency modes of the harmonic chain of finite size $N$. Expressions~(\ref{for}) and (\ref{for1}) determine how the heat flux scales with the system size, and with the addition of Eq.~(\ref{ther}) the size scaling law of the thermal conductivity is obtained as well. Indeed, in the case when $W(\mu)\sim \mu^\beta$ as $\mu \rightarrow 0$, Eq.~(\ref{scale}) is obtained. However, we must keep in mind that Eqs.~(\ref{for1}) may not be valid if the inverse localization length is zero for a frequency $\omega_c \ne 0$. In such  situation, the vibrational modes around $\omega_c$ contribute to the heat flux as well. This occurs for example in the random dimer model \cite{BG20}.

Using Eq.~(\ref{for}) and Eq.~(\ref{powerT}) for $\alpha=2$, the following transcendental equation for $N_e$ is obtained 
\begin{equation}
N^2_e\left[\ln N -\ln(\pi N_e) \right]=\frac{8M^2}{\pi^2\text{var}[m_n]}\frac{\zeta(2)[\zeta(2)-\zeta(3)]}{\zeta(3)}N.
\label{trans}
\end{equation} 
This equation can only be solved numerically. Indeed, one can not neglect the term $\ln(\pi N_e)$ even for $\ln N\gg1$, we prove this by the contradiction method. Assume that $\ln(\pi N_e)$ can be neglected for huge system sizes, then the following closed solution is obtained $N_e \sim \left(N/\ln N\right)^{1/2}$; this implies that the condition $\ln N\gg \ln N_e$ is no longer valid, but this corresponds to our initial hypothesis. Therefore, the hypothesis is wrong.

We have investigated numerically the solutions of Eq.~(\ref{trans}) which has two solutions: the first one scales roughly as $N^{1/2}$, while the largest solution scales roughly as $N$. However, there are some corrections to this scaling laws which can only be determined by solving numerically this equation. In addition, the smallest solution is the one that represents the true physical solution since it is clear that $N_e$ can not scale roughly as $N$ for $\alpha=2$. Thus, the smallest solution to Eq.~(\ref{trans}) together with Eq.~(\ref{for1}) represent a non closed solution for the asymptotic scaling law of the thermal conductivity.

We can obtain an estimation of how the thermal conductivity scales for $\alpha<1$ if one uses expression (\ref{am}) which was obtained only by numerical simulations. Indeed, the $n$th low frequency state is extended if its localization length $L_{\text{loc}}$ is greater than the system size
\begin{eqnarray}
N \lesssim  L_{\text{loc}}\left(\frac{\omega_{\text{max}}\pi n}{N} \right), 
\label{sim2} 
\end{eqnarray}
where it is assumed that the $n$th frequency is practically the same than the corresponding one of an ordered harmonic chain, then the dispersion relation is linearized and the latter equation is obtained. Therefore, the number of low frequency vibrational modes $N_e$ that contributes to the heat conduction is determined by replacing $n$ by $N_e$ and  the symbol $\lesssim$ by $\approx$ in Eq.~(\ref{sim2}). Thus, with the use of Eq.~(\ref{sim2}), the number of low frequency vibrational modes scales as 
\begin{eqnarray}
N_e\sim N^{(1-\alpha)/(2-\alpha)} \qquad \qquad \text{for} \ \alpha<1.
\end{eqnarray}
When this expression is replaced in Eq.~(\ref{for1}) and if the definition of thermal conductivity (\ref{ther}) is taken into account, the following expressions for the asymptotic scaling law of the thermal conductivity is obtained for $\alpha<1$
\begin{eqnarray}
    \kappa\sim
         \left\{
                \begin{array}{ll}
                  N^{-(\alpha+1)/(2-\alpha)} & \ \mbox{for} \ \ \text{fixed BC} \\
                  N^{(1-\alpha)/(2-\alpha)} & \ \mbox{for} \ \ \text{free BC}
                \end{array}
         \right. ,
\label{km}
\end{eqnarray}
therefore, a decreasing (increasing) thermal conductivity with the system size is obtained for fixed BC (free BC). Notice that as $\alpha \rightarrow 0$, the result for the uncorrelated disorder case is obtained, which is somehow an unexpected result, however, we must keep in mind that the Lyapunov exponent $\lambda$ goes to zero as $\alpha$ is decreased, which means that the scaling law (\ref{km}) is reached for larger values of $N$ as $\alpha$ decreases. This means that the size of the region for which $L_{\text{loc}}\gg N$ is incremented as $\alpha$ is decreased. In this region, the thermal conductivity is ballistic and $\kappa \sim N$. Thus, when $\alpha \rightarrow 0$, the size of this region becomes infinite, and the expected result $\kappa \sim N$ is obtained regardless on the boundary conditions.

\begin{figure}[!t]
\includegraphics[width=8.8cm,height=6.4cm]{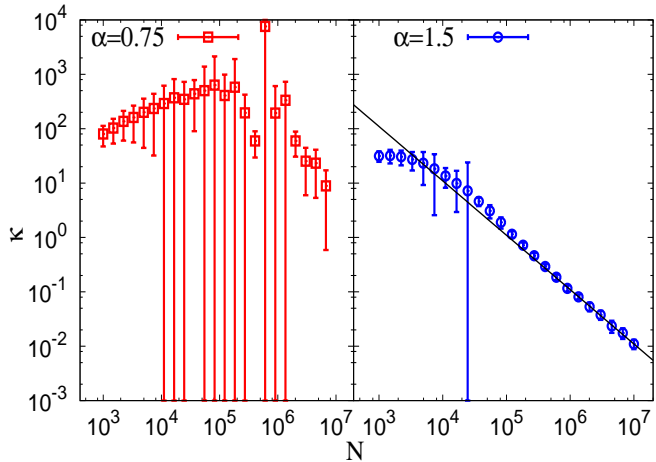}
\caption{(Color online) Thermal conductivity $\kappa$ vs system size $N$ for fixed boundary conditions. The points represent the mean value of the conductivity, obtained after averaging over 100 disorder realizations (except for the data between $N=10^6$ and $N=10^7$ where only $10$ disorder realizations were considered), whereas the continuous line in the right panel corresponds to the best fit of the asymptotic data to the function $f(N)=aN^b$, with $a=109871\pm 1948$ and $b=1.01 \pm 0.03$. In both panels $M=1$, $m=0.75$, $\gamma=1$, and $\triangle T=1$.}
\label{Fig2F}
\end{figure}

\begin{figure}[!t]
\includegraphics[width=8.8cm,height=6.4cm]{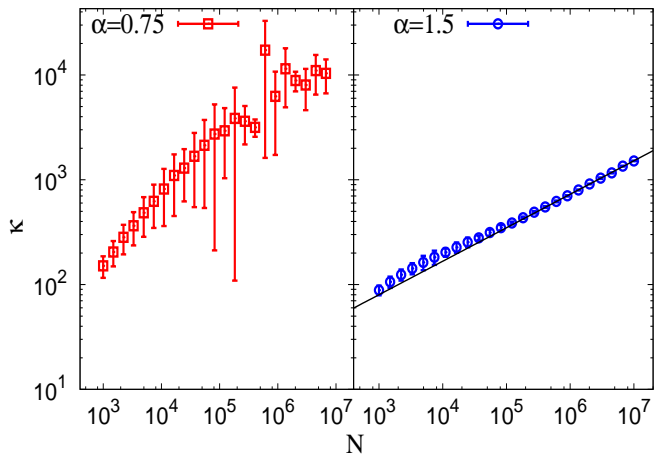}
\caption{(Color online) Thermal conductivity $\kappa$ vs system size $N$ for free boundary conditions. The points represent the mean value of the conductivity, whereas the continuous line in the right panel corresponds to the best fit of the asymptotic data to the function $f(N)=aN^b$, with $a=8.8\pm 0.6$ and $b=0.32\pm 0.01$. Same parameters and number of disorder realizations as in Fig.~\ref{Fig2F}.}
\label{Fig3L}
\end{figure}

To validate the above results numerically, we use the integral method introduced in Ref.~\cite{D01} which is only valid for harmonic chains. The results are shown in Figs.~\ref{Fig2F} and \ref{Fig3L}, where for the case of $\alpha<1$ (see left panels), thermal conductivity shows strong fluctuations even for system sizes of length $N=10^7$. Unfortunately, this does not allow us to verify Eq.~(\ref{km}) since larger system sizes are needed to reduce these fluctuations. The origin of these strong fluctuations is due to the phenomenon of L\'evy flights which becomes negligible for $N\gg1$; the smaller the exponent $\alpha$, the larger the system size $N$ needed to reduce the fluctuations of the thermal conductivity. Therefore, we could  corroborate the size scaling law predicted by Eq.~(\ref{scaleT}) only for $\alpha\gtrsim1.2$. In Figs.~\ref{Fig2F} and \ref{Fig3L} (right panels), only the case $\alpha=1.5$ is shown.

It is worthwhile to mention that our results (localization length and size scaling of thermal conductivity) can only be applied to mass disorder since they are derived using the analogy that exists between the disordered harmonic chain model and the 1D Anderson model with diagonal disorder. When random spring constants are considered, non-diagonal disorder appears in Eq. (\ref{Fou}); we refer to the reader to references \cite{AOI18,AABOI19} where analytical solutions are obtained for the size scaling behavior of the thermal conductivity (in different disordered regimes) for a chain whose spring constants are distributed randomly with a L\'evy--type distribution or a box distribution.
\section{Conclusions}

We have analyzed the localization  and thermal properties of a harmonic chain with L\'evy--type disorder which is attached at its ends to two Langevin heat baths. The complete analytical solution for the localization length was found when $\alpha>1$. In addition, we showed that formulas previously derived for the localization length have a wider range of validity. When $1<\alpha<2$, the inverse localization length scales as $L^{-1}_{\text{loc}}(\omega)\sim \omega^{\alpha}$, whereas for $\alpha>2$, the uncorrelated disorder case is recovered ($L^{-1}_{\text{loc}}(\omega)\sim \omega^{\alpha}$). For $\alpha=2$, $L^{-1}_{\text{loc}}(\omega)\sim\omega^2 \ln \omega$. When $\alpha\le1$, our analytical results can not be applied, so for this situation, we rely on numerical simulations to get insight about the localization length. In particular, for $\alpha<1$, $L^{-1}_{\text{loc}}(\omega)\sim \omega^{2-\alpha}$; here we again stress that the inverse localization length is finite, contrary to the common belief where it was assumed that $L^{-1}_{\text{loc}}$ goes to zero in the thermodynamic limit for $\alpha<1$.

The above results allowed us to give an analytical estimation of how the asymptotic thermal conductivity scales with the system size: The results depend on boundary conditions as usual, and they are given by Eqs.~(\ref{scaleT}), (\ref{scaleT1}) and (\ref{km}). Unfortunately, we were not able to corroborate numerically the case $\alpha<1$ due to the phenomenon of L\'evy flights which produces strong fluctuations of the thermal conductivity for the system sizes analyzed in this work.

A remarkable situation occurs when  $\alpha=2$, since  due to the logarithmic correction of the localization length, the asymptotic scaling law of the thermal conductivity acquires a non-closed form which represents an unusual situation because the typical scaling law of the thermal conductivity for 1D systems is $\kappa \sim N^{\alpha_1}$, where the value of the exponent $\alpha_1$ depends on the particular one-dimensional model \cite{D01,LXXZZL12}. Moreover, this logarithmic correction may be tested experimentally by measuring  the electromagnetic transmission through one-dimensional photonic heterostructures whose random layer thicknesses follow a long-tailed L\'evy-type distribution since now there exist experimental devices where the anomalous localization of waves can be induced in a controllable manner \cite{FMCCSG14}. The same phenomenon, may have some implications to the transmission $T$ of light in L\'evy glasses as a function of their thickness $L$ since it may appear some corrections to the typical diffusive behavior for which $T \sim L^2$, $L\gg 1$, however, this corrections will be not easy to observe, and an extension of the model analyzed in this work to higher dimensions is needed to provided a firmer theoretical background to the phenomenon of light in L\'evy glasses.

\begin{acknowledgments}
J.A.M.-B. thanks support from 
FAPESP (Grant No. 2019/06931-2), Brazil, and
VIEP-BUAP (Grant No. 100405811-VIEP2019) and
Fondo Institucional PIFCA (Grant No. BUAP-CA-169), Mexico. 
\end{acknowledgments}

\end{document}